\documentclass[
 ,final            
 ,numberedheadings 
  ,nomathfonts                 
  ]
  {aipproc}

\layoutstyle{8x11single}

\newcommand{\beq}{\begin{equation*}}
\newcommand{\eeq}{\end{equation*}}
\newcommand{\be}{\begin{equation}}
\newcommand{\ee}{\end{equation}}
\newcommand{\beqa}{\begin{eqnarray}}
\newcommand{\eeqa}{\end{eqnarray}}
\newcommand{\bea}{\begin{eqnarray}}
\newcommand{\eea}{\end{eqnarray}}
\newcommand{\bra}{\langle}
\newcommand{\ket}{\rangle}

\newcommand{\s}{\sigma}
\newcommand{\w}{\omega}

\newcommand{\emdash}{\hspace{1pt}---\hspace{1pt}}

\newcommand{\putat}[3]{\begin{picture}(0,0)(0,0)\put(#1,#2){#3}\end{picture}}

\begin{document}

\title[The Quark-Meson Coupling model as a description of dense
  matter]{The Quark-Meson Coupling model as a\\ description of dense matter}



\author{J.~D.~Carroll}{ address={Centre for the Subatomic Structure of
    Matter (CSSM), Department of Physics,\\ University of Adelaide, SA
    5005, Australia} }

\begin{abstract}
Quantum Hadrodynamics provides a useful framework for investigating
dense matter, yet it breaks down easily when strangeness carrying
baryons are introduced into the calculations, as the baryon effective
masses become negative due to large meson field potentials. The
Quark-Meson Coupling model overcomes this issue by incorporating the
quark structure of the nucleon, thus allowing for a feedback between
the the nuclei and the interaction with the meson fields. With the
inclusion of this feature, QMC provides a successful description of
finite nuclei and nuclear matter. We present the latest
parameterization of QMC and discuss the predictions for dense nuclear
matter and `neutron' stars.
\end{abstract}

\maketitle


\section{INTRODUCTION}

 The Quark-Meson Coupling model
 (QMC)~\cite{Guichon:1995ue,QMC2007,QMC2008,JDC2009,JDCthesis,JDC60fest}
 has proven to be an extremely useful tool for calculating the
 properties of matter at various density scales\emdash from finite
 nuclei~\cite{QMC2008} to neutron stars~\cite{JDC2009}\emdash where
 the global properties of these objects can be accurately predicted
 and compared with experiment in an effort to refine the model
 parameters.\par

 While other models for these systems have been investigated by many
 others~\cite{Serot:1984ey,Pieper:2007ax,Wang:2005vg}, QMC uniquely
 distinguishes itself as a relativistic description at the quark
 level, constrained only via the quark-structure of baryons and a few
 baryon-meson coupling constants (for a thorough description,
 see~\cite{JDCthesis}), fit to experimental data.\par

\section{QMC}\label{sec:QMC}

 In order to calculate the properties of finite nuclei (including
 hypernuclei) and neutron stars, we require a parameterization of the
 baryon effective masses, $M^*_B$ which is determined
 self-consistently using the QMC model, and which has a quadratic
 dependence on the scalar mean-field. This quadratic dependence is a
 distinguishing feature of QMC as compared to Quantum Hadrodynamics
 (QHD) which models a linear dependence of the effective mass on the
 scalar mean-field~\cite{Serot:1984ey}; a feature which leads to
 negative effective masses at large baryon densities.\par

\begin{figure}[!t]
\begin{tabular}{lr}
\begin{minipage}[c]{0.45\textwidth}
  \hspace{-8mm}
  \includegraphics[angle=90,width=1.1\textwidth]{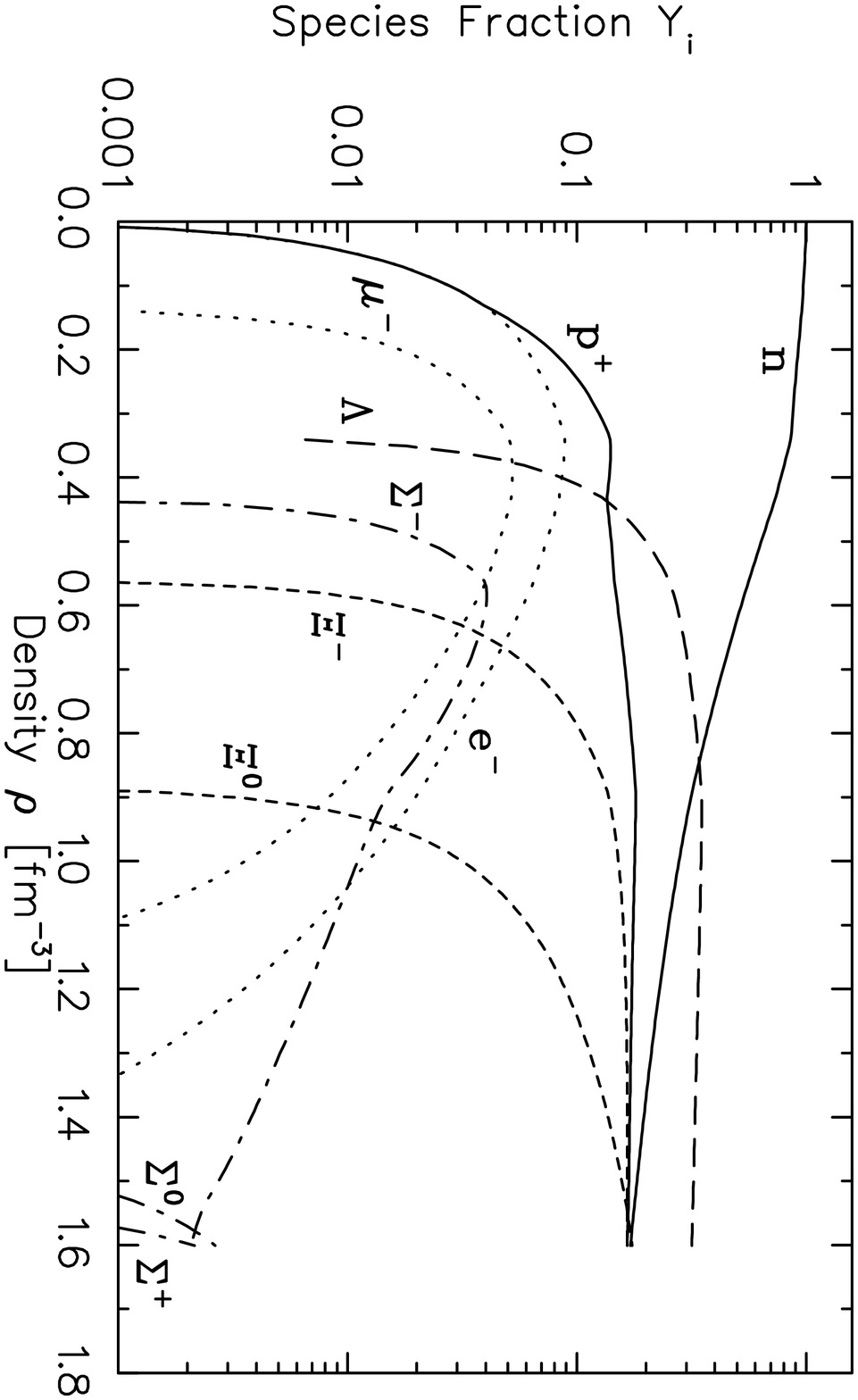}
\end{minipage}
&
\begin{minipage}[c]{0.45\textwidth}
  \includegraphics[angle=90,width=1.1\textwidth]{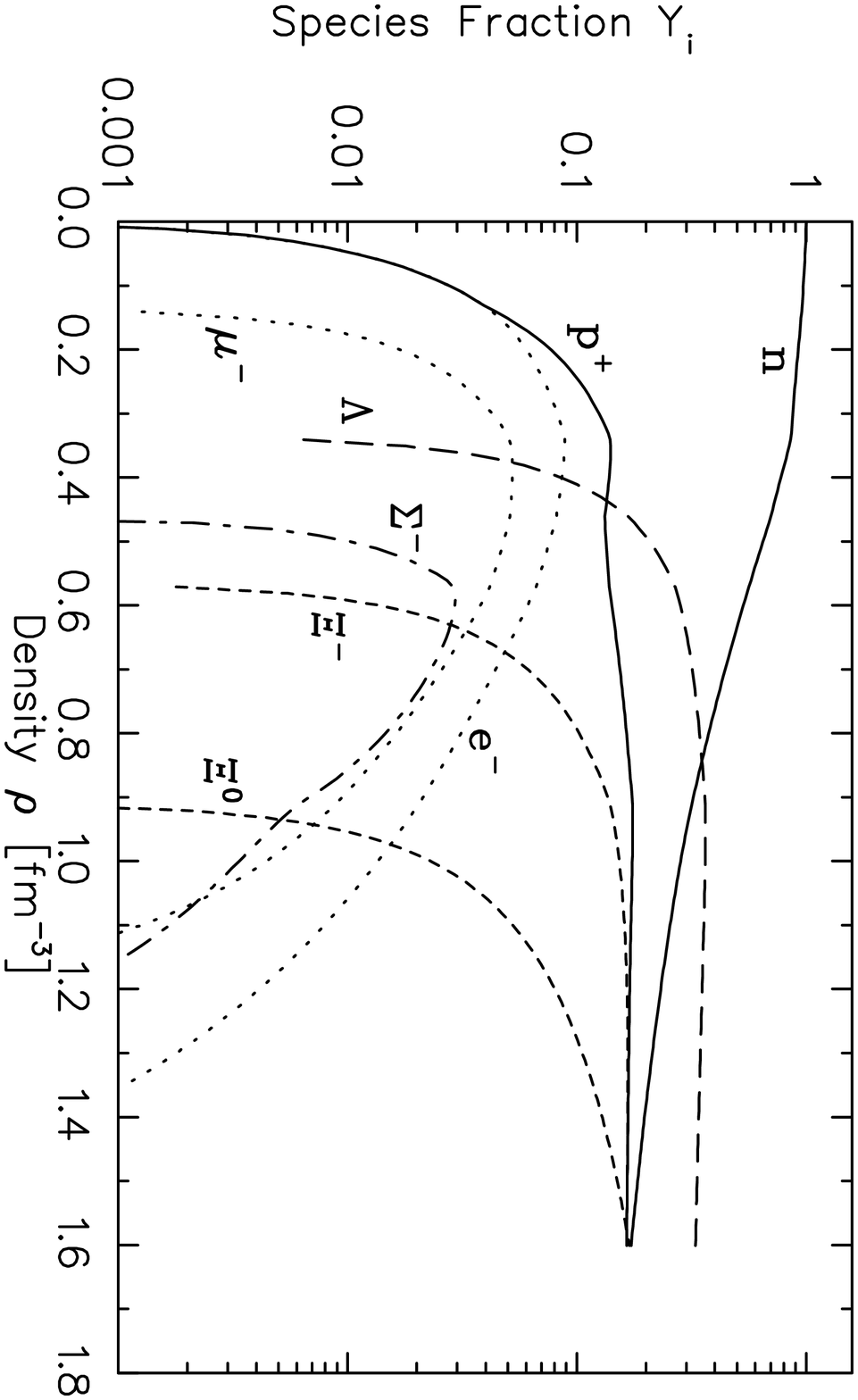}
\end{minipage}
\end{tabular}
  \caption{Species fractions $Y_i$ for hyperonic QMC hadronic matter
    calculated using the 2007 ((a),~from~\cite{QMC2007}), and 2008
    ((b),~from~\cite{QMC2008}) parameterizations of the effective
    masses. Note the suppression of the $\Sigma$ baryon densities
    predicted using the 2008
    parameterization. \protect\label{fig:SpecFrac_comp}} 
  \putat{-260}{60}{(a)}
  \putat{-17}{60}{(b)}
\end{figure}

 With a parameterization of $M^*_B$, we are able to calculate the
 properties of infinite nuclear matter, in which baryons are in
 beta-equilibrium with leptons, including attractive and repulsive
 potential contributions from scalar $\s$, and vector $\w$ and $\rho$
 mesons. For example, using the parameterization found in
 Ref.~\cite{QMC2007} we calculate the properties of infinite matter
 under the constraints of global charge neutrality and conserved
 baryon density, as shown in Figure~\eqref{fig:SpecFrac_comp}(a). The
 couplings constants $g_{\s}$, $g_{\w B}$, and $g_{\rho}$ of the
 baryons $B$ to the various mesons are determined by the standard
 procedure of fitting the properties of saturated nuclear matter.\par

 We note that the fractional density of $\Sigma^-$ baryons decreases
 above a certain density, while that of the remaining baryons tend
 toward a common value at large densities. We also note that the
 threshold density for $\Xi$ hyperons to appear is well below that
 of the $\Sigma^0$ and $\Sigma^+$ hyperons.\par

 The parameterization of the baryon effective masses has been refined
 over time to better reflect the properties of hypernuclei. Most
 recently in particular in 2008, by the self-consistent inclusion of
 one-gluon-exchange hyperfine terms that lead to the $N$-$\Delta$ and
 $\Sigma$-$\Lambda$ mass splittings in free-space. A consequence of
 this is that the $\Sigma$ hypernuclei are unbound in this model,
 which is consistent with the lack of evidence from the experimental
 searches for these. The parameterization is given in terms of the
 scalar mean-field $\bra\s\ket$ by \par

\begin{eqnarray} \label{eq:MstarsinQMC}
M^*_{N} & = & M_{N}-g_{\s N}\bra\s\ket\nonumber \\\nonumber
& &+\left[0.002143+0.10562R_{N}^{\rm free}-0.01791\left(R_{N}^{\rm free}\right)^{2}\right]
\left(g_{\s N}\bra\s\ket\right)^{2},
\label{eq:A18}\\
M^*_{\Lambda} & = & M_{\Lambda}-\left[0.6672+0.04638R_{N}^{\rm free}-0.0022\left(R_{N}^{\rm free}\right)^{2}\right]g_{\s N}\bra\s\ket
\nonumber \\ \nonumber
 &  & +\left[0.00146+0.0691R_{N}^{\rm free}-0.00862\left(R_{N}^{\rm free}\right)^{2}
\right]\left(g_{\s N}\bra\s\ket\right)^{2},
\\ \label{eq:A19usethis}
M^*_{\Sigma} & = & M_{\Sigma}-\left[0.6653-0.08244R_{N}^{\rm free}+0.00193\left(R_{N}^{\rm free}\right)^{2}\right]g_{\s N}\bra\s\ket
 \\\nonumber
 &  & +\left[0.00064+0.07869R_{N}^{\rm free}-0.0179\left(R_{N}^{\rm free}\right)^{2}
\right]\left(g_{\s N}\bra\s\ket\right)^{2},
\label{eq:A21}\\
M^*_{\Xi} & = & M_{\Xi}-\left[0.3331+0.00985R_{N}^{\rm free}-0.00287\left(R_{N}^{\rm free}\right)^{2}\right]g_{\s N}\bra\s\ket
\nonumber \\\nonumber
 &  & +\left[-0.0032+0.0388R_{N}^{\rm free}-0.0054\left(R_{N}^{\rm free}\right)^{2}
\right]\left(g_{\s N}\bra\s\ket \right)^{2}\, ,
\label{eq:A22}
\end{eqnarray}

 where the value $R_{N}^{\rm free}=0.8~{\rm fm}$ is used (and the
 dependence on this value has been investigated and found to be small
 in Ref.~\cite{Guichon:2006er}).\par

 The relative densities of baryons and leptons in beta-equilibrium for
 the case of infinite nuclear matter for this new (2008)
 parameterization are shown in Figure~\eqref{fig:SpecFrac_comp}(b). We
 note that the most striking difference in comparison to the
 calculation using the 2007 parameterization\emdash which does
 parameterize the additional hyperfine interactions\emdash is that the
 contribution of the $\Sigma$ hyperons has been reduced.\par

\section{Neutron Stars}\label{sec:TOV}

\begin{figure}[pbt]
  \includegraphics[angle=90,width=0.6\textwidth]{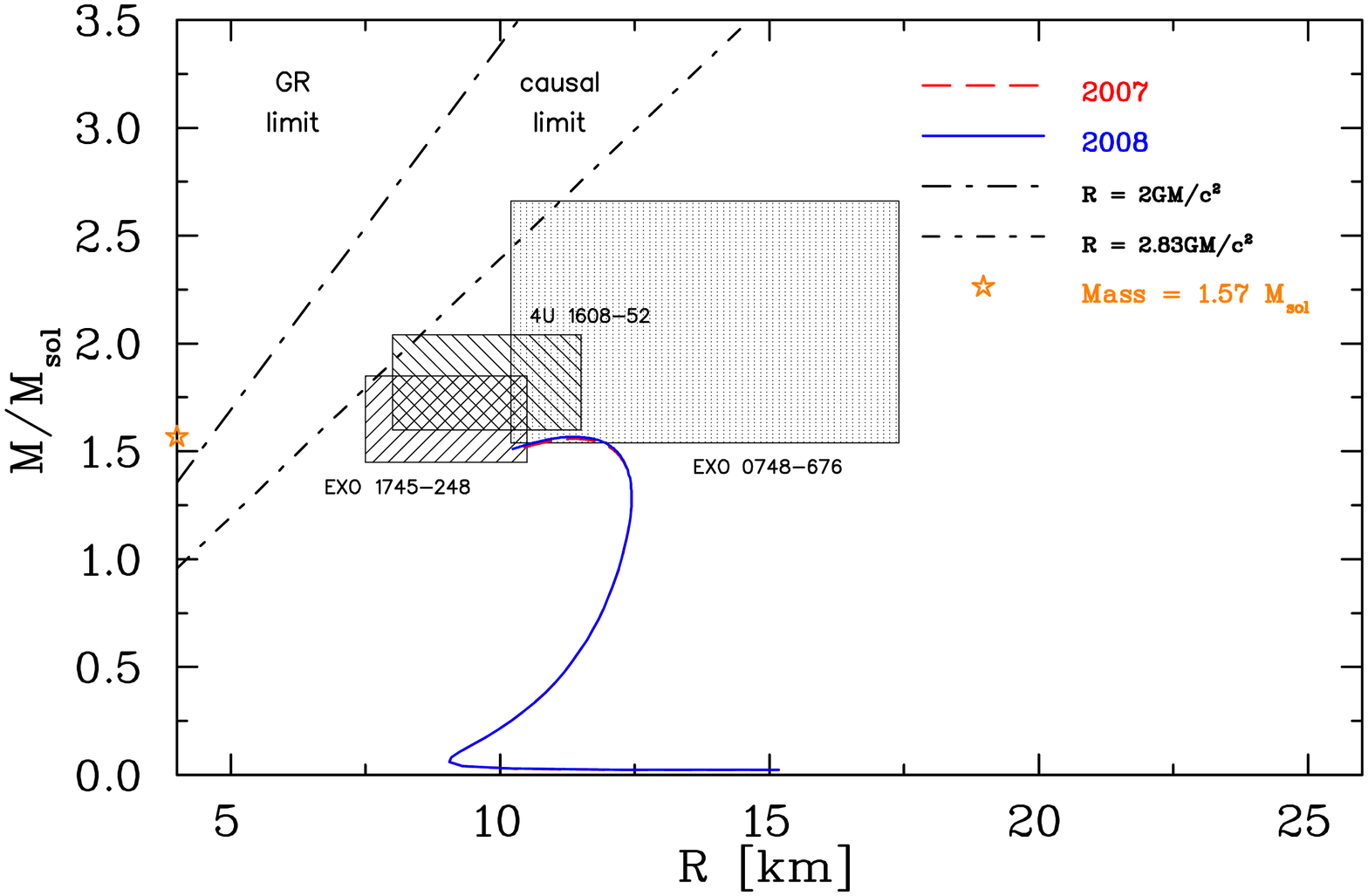}
  \caption{(Color online) Predicted total stellar mass and radius
    solutions of the TOV equation for hyperonic QMC using the 2007 and
    2008 parameterizations of the effective mass. Also shown are the
    data points from Refs.~\cite{Ozel:2006bv} (EXO 0748-676),
    \cite{Guver:2008gc} (EXO 1745-248), and \cite{Ozel:2008kb} (4U
    1608-52), as well as the maximum mass predicted from either data
    set (star). Inset: Separation of the two data sets at the maximum stellar
    mass.  \protect\label{fig:TOV_comp}} \putat{-105}{34}{
    \includegraphics[angle=90,width=0.19\textwidth]{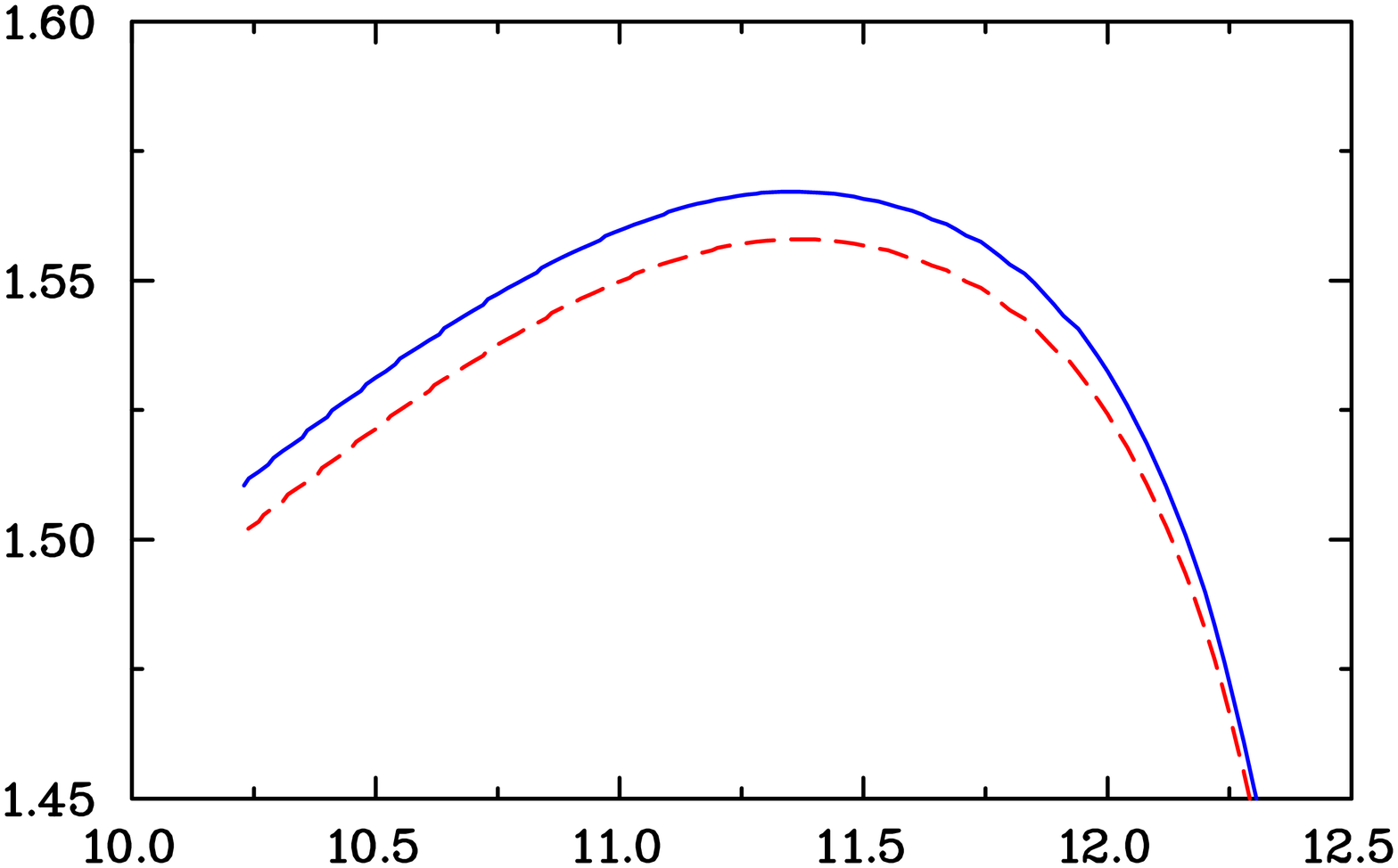}}
\end{figure}

 Calculating the mass and radius of a neutron star modelled with QMC
 involves solving the Tolman-Oppenheimer-Volkoff (TOV) equation 
\be \label{eq:TOVdpdr}
\frac{dP}{dR} = 
-\frac{G\left(P + {\cal E}\right)\left(M(R)+4\pi R^3P\right)}{R(R-2GM(R))},
\ee
 for a given Equation of State (EoS), in this case an EoS calculated
 using QMC (see Refs.~\cite{JDCthesis,JDC60fest} for details). The TOV
 solutions for a range of central densities are shown in
 Figure~\eqref{fig:TOV_comp} for the 2007 and 2008 parameterizations
 of the effective masses. We note that the choice of parameterization
 does not significantly affect the solutions to the TOV equation.\par

 We further note that the maximum predicted stellar mass calculated
 using either parameterization of the effective baryon mass does not
 exceed 1.57~$M_\odot$, which can be attributed to the softness of the
 EoS due to the large number of baryons present\emdash at large
 densities, at least six of the octet baryons posess nontrivial
 fractional densities. This is in conflict with the observational
 evidence of larger mass neutron stars, though we caution using a
 direct comparison between these predictions (which correspond to
 static, spherically symmetric, non-rotating objects) and physical
 neutron stars which may not satisfy such approximations.\par

 Future work will include investigating the effects of rotation on the
 stellar solutions, and inclusion of Fock (exchange) terms to the
 baryon self-energies, in an effort to increase the accuracy of our
 description of dense nuclear matter.


\begin{theacknowledgments}
  This research was supported in part by DOE contract
  DE-AC05-06OR23177 (under which Jefferson Science Associates, LLC,
  operates Jefferson Lab), and in part by the Australian Research
  Council. The author would like to thank A.~W.~Thomas for his
  guidance and support, as well as D.~B.~Leinweber and A.~G.~Williams
  for their helpful discussions.
\end{theacknowledgments}



\bibliographystyle{aipproc}   




\end{document}